\documentclass[10pt,conference]{IEEEtran}
\usepackage{graphicx}
\usepackage{cite}
\usepackage{amsthm}
\usepackage{balance}

\usepackage{fancyhdr}

\usepackage{xcolor} 
\usepackage{algorithm2e} 
\usepackage{braket} 
\usepackage{url}

\title{Hybrid Meta-Solving for Practical~Quantum~Computing}

\author{
    \IEEEauthorblockN{Domenik Eichhorn}
    \IEEEauthorblockA{
        Karlsruhe Institute of Technology\\
        Karlsruhe, Germany\\
        domenik.eichhorn@kit.edu
        }
    \and
    \IEEEauthorblockN{Maximilian Schweikart}
    \IEEEauthorblockA{
        Karlsruhe Institute of Technology\\
        Karlsruhe, Germany\\
        maximilian.schweikart@student.kit.edu
        }
    \and
    \IEEEauthorblockN{Nick Poser}
    \IEEEauthorblockA{
        Karlsruhe Institute of Technology\\
        Karlsruhe, Germany\\
        nick.poser@student.kit.edu
        }
    \and
    \IEEEauthorblockN{Frederik Fiand}
    \IEEEauthorblockA{
        GAMS Software GmbH\\
        Frechen, Germany\\
        ffiand@gams.com
        }
    \and
        \IEEEauthorblockN{Benedikt Poggel}
    \IEEEauthorblockA{
        Fraunhofer Institute for Cognitive Systems IKS\\
        Munich, Germany\\
        benedikt.poggel@iks.fraunhofer.de
        }
    \and
    \IEEEauthorblockN{Jeanette Miriam Lorenz}
    \IEEEauthorblockA{
        Fraunhofer Institute for Cognitive Systems IKS\\
        Munich, Germany\\
        jeanette.miriam.lorenz@iks.fraunhofer.de
        }
}

\fancypagestyle{specialfooter}{%
  \fancyhf{}
  
  \fancyfoot[R]{ \noindent\fbox{%
    \parbox{\textwidth}{ \centering
        {\footnotesize This work has been submitted to the IEEE for possible publication. \\ 
        Copyright may be transferred without notice, after which this version may no longer be accessible.}
        }
    }}
}
\begin{document}

\maketitle

\thispagestyle{specialfooter}

\begin{abstract}
The advent of quantum algorithms has initiated a discourse on the potential for quantum speedups for optimization problems. 
However, several factors still hinder a practical realization of the potential benefits.
These include the lack of advanced, error-free quantum hardware, the absence of accessible software stacks for seamless integration and interaction, and the lack of methods that allow us to leverage the theoretical advantages to real-world use cases.
This paper works towards the creation of an accessible hybrid software stack for solving optimization problems, aiming to create a fundamental platform that can utilize quantum technologies to enhance the solving process.
We introduce a novel approach that we call Hybrid Meta-Solving, which combines classical and quantum optimization techniques to create customizable and extensible hybrid solvers. 
We decompose mathematical problems into multiple sub-problems that can be solved by classical or quantum solvers, and propose techniques to semi-automatically build the best solver for a given problem. 
Implemented in our ProvideQ toolbox prototype, Meta-Solving provides interactive workflows for accessing quantum computing capabilities.
Our evaluation demonstrates the applicability of Meta-Solving in industrial use cases.
It shows that we can reuse state-of-the-art classical algorithms and extend them with quantum computing techniques. Our approach is designed to be at least as efficient as state-of-the-art classical techniques, while having the potential to outperform them if future advances in the quantum domain are made.
\end{abstract}


\section{Introduction}
\label{sec:introduction}
Quantum algorithms have demonstrated theoretical advantages over their classical counterparts in addressing problems such as unstructured database search~\cite{grover1996fast}, factorization~\cite{shor1999polynomial}, and testing whether a function is constant~\cite{deutsch1992rapid}. 
These theoretical advantages have prompted a discussion of potential applications of quantum technologies to the optimization domain, with the objective of retrieving practical quantum speedups and creating more efficient solvers.
Nevertheless, we are currently in the early stages of quantum computing, where the practical quantum advantages for optimization problems have yet to be realized~\cite{stilck2021limitations, bharti2022noisy}. 
There are several factors currently preventing us from achieving quantum supremacy, a major one being the availability of scalable quantum hardware with large numbers of qubits with high connectivity and efficient error correction.

However, the availability of advanced quantum hardware does not guarantee the development of superior solvers. 
We have yet to identify methods for translating the theoretical speedups into practical applications. To this end, we must create software stacks that facilitate the integration of quantum solutions into broader computational pipelines, where they can operate in conjunction with classical computers in an efficient and effective manner.
Moreover, it is necessary to investigate how the theoretical advantages of quantum computing can be applied in actual computational pipelines, where information between classical and quantum computers must be transferred continuously.
Currently, encoding information on quantum computers requires extensive transformation techniques. 
For instance, when creating oracles to apply Grover's algorithm~\cite{grover1996fast}, or when transforming constrained algorithms into Quadratic Unconstrained Binary Optimization (QUBO) Problems to apply quantum approximation algorithms~\cite{farhi2014quantum}.

Next, quantum computing must be made more accessible to the general public.
Vendors of quantum solutions require their users to utilize their frameworks in a manner that is opaque to the user, limiting their ability to adapt the framework to diverse real-world problems~\cite{osaba2024hybrid}. 
In other instances, users are provided with only basic programming kits and frameworks, such as Qiskit~\cite{Qiskit}, Pennylane~\cite{bergholm2022pennylane}, or Qrisp~\cite{seidel2022qrisp}.
These frameworks require users to possess advanced expertise and to implement the core functionality themselves.
Both of these options are suboptimal, as users should not be forced to identify opportunities and implement quantum applications themselves. 
Rather, they should have the ability to customize quantum application pipelines to optimize their performance and meet their custom needs. Ultimately, an abstraction layer that covers both the classical and quantum parts of the computation is needed.

This paper works towards the creation of an accessible hybrid software stack for solving optimization problems, aiming to create a fundamental platform that can utilize quantum technologies to enhance the solving process.
We introduce a concept called Hybrid Meta-Solving, which combines the advantages of classical and quantum optimization in hybrid solution strategies to create new, powerful ways to solve well-known mathematical problems. 
Meta-Solving describes the decomposition of a mathematical problem into multiple sub-problems, each of which can be solved by a selection of solvers. Using expert knowledge, empirical data, and established heuristics, we can compare potential classical and quantum solvers for a subroutine and find the best solver for the given problem. 
This paper outlines the fundamental concepts of Meta-Solving and illustrates how these concepts can be utilized to create interactive, semi-automated workflows. 
We explain how users can utilize those workflows to exploit the potential of quantum computing and find efficient solutions for given algorithmic problems. 
A first prototype implementing the fundamentals of Meta-Solving is available in our ProvideQ toolbox~\cite{eichhorn2023providing}.
Our evaluation demonstrates that our Meta-Solving concept is applicable to realistic problems and reaches at least the same performance as classical state-of-the-art approaches. 
While we are not yet able to reach actual quantum speedups, we show how a fundamental platform that integrates classical and quantum techniques can be created.


\section{Background and Related Work}
\label{sec:background}
This section briefly introduces the background to quantum computing and presents state-of-the-art approaches and existing work related to our Meta-Solving concept. 

\subsection{Current-era Quantum Computing}


Today we are in what is known as the Noisy Intermediate-Scale Quantum (NISQ)~\cite{preskill_quantum_2018} era. 
Medium-scale quantum computers with a few hundred qubits are available and can be programmed using a gate-based programming model.
However, the hardware is still noisy, and it requires expensive error-mitigation measures to produce reasonable results even for very small problems~\cite{abughanem2023nisq}.
A plethora of quantum algorithms are currently being studied on small, error-prone quantum computers or simulators, as well as through theoretical means. 
Algorithms such as Grover~\cite{grover1996fast}, Shor~\cite{shor1999polynomial}, and Deutsch-Jozsa~\cite{deutsch1992rapid} were designed even before the first quantum computers became available. 
These algorithms provide theoretically proven advantages, but we are currently unable to leverage them in practice due to a number of factors, including the fact that they were designed for fault-tolerant quantum computers, which are not yet available. 
To make quantum computing viable in the near future, NISQ-tailored quantum algorithms such as the Quantum Approximate Optimization Algorithm (QAOA)~\cite{farhi2014quantum} and the Variational Quantum Eigensolver (VQE)~\cite{peruzzo2014variational} have been developed. 
However, it has not yet been demonstrated that the NISQ-tailored algorithms can provide actual speedups. 

\subsection{Quantum Computing Platforms for Optimization}

With the constant improvement of the capacities of actual quantum hardware and the new possibilities for algorithms executed on it, various endeavours have started working on an abstraction layer that relieves the end user from deciding between the numerous options. Formally, one can integrate these options in a modular decision tree with a set of options then forming a so-called Solution Path, and recommend Solution Paths based on various metrics and characteristics of the application~\cite{poggel_recommending_2023}. Finding and evaluating good Solution Paths for application problems like vehicle routing is hard, however, and requires extensive domain knowledge along with hardware improvements and computational tests~\cite{palackal_quantum-assisted_2023}. Hybrid solvers are in development, e.g., by Quantagonia~\cite{quantagonia_hybridsolver} or D-Wave~\cite{d-wave_developers_d-wave_2020}, though focusing mostly on annealing methods for now due to their farther maturity. A thorough description and benchmark is found in~\cite{osaba2024hybrid}. With the PlanQK platform~\cite{planqk}, a first hardware-agnostic platform and vision on how end users can approach solving various application cases with quantum-enhanced algorithms exists. The efforts for abstraction go beyond optimization and similarly extend to Quantum Machine Learning~\cite{klau_autoqml_2023}. 
Our work focuses on the integration of quantum computing and its existing platforms into classical optimization techniques. We combine the best of both worlds with the goal of building a new platform that gives users the tools they need to take advantage of quantum computing and build efficient hybrid solvers tailored to their needs. 

\subsection{Classical Optimization and Polylithic Modeling}

In the domain of classical mathematical optimization, specialized solvers have been developed to address particular problems with high efficiency, exemplified by the TSP solver Concorde~\cite{concorde}. Additionally, optimization solvers that cater to broader problem classes, such as Linear Programming (LP), Mixed-Integer Linear Programming (MILP), Nonlinear Programming (NLP), and Mixed-Integer Nonlinear Programming (MINLP), exhibit considerable computational power. These solvers have undergone continuous refinement resulting in remarkable speedup over several decades~\cite{Koch2022, bussieck2011} and are presently employed to tackle complex, real-world challenges.

However, as solvers become more efficient, users strive to build more accurate models, thereby increasing their complexity.
Despite the advancements in optimization solver technology, a number of practical optimization challenges remain that are not adequately addressed by current state-of-the-art solutions. 
In response to this gap, methodologies that decompose a complex, "monolithic" problem into a series of simpler, more manageable subproblems have gained prominence. 
Termed "polylithic" modeling and solution approaches, this strategy entails the development of customized methods that incorporate multiple models and/or algorithmic components~\cite{Kallrath2011}. Here, the solution derived from one model serves as the input for another. 
Notable examples of such polylithic approaches include decomposition techniques (e.g. Benders~\cite{benders2005partitioning} and Dantzig-Wolfe~\cite{dantzig1961decomposition} decomposition), advanced MILP and MINLP solvers that integrate presolve strategies with the sequential resolution of subproblems (frequently employing various external sub-solvers) within a Branch and Cut framework, and hybrid methods that integrate constructive heuristics and local search improvement strategies with exact Mathematical Programming algorithms.
While polylithic modeling is not inherently dependent on any specific software, algebraic modeling languages such as the General Algebraic Modeling System (GAMS), have demonstrated significant utility in facilitating the implementation of these sophisticated approaches~\cite{Kallrath2011}.

In our work, we draw inspiration from well-established polylithic approaches in classical optimization and extend them with quantum computing techniques. 
We actively reuse openly available state-of-the-art solvers and decompositions to maximize the efficiency of our approach. 
Furthermore, we bundle the existing state-of-the-art into a user-friendly toolbox to enable easy reusability, and extensibility. 


\section{The Meta-Solving Concept}
\label{sec:concept1}

Following, we explain the general idea and core concepts of Meta-Solving.
Meta-Solving combines classical optimization principles with quantum computing algorithms to create new, powerful solvers. 
We facilitate the access to quantum computing solutions by introducing a concept that reuses the advantages of classical polylithic approaches and extends them with quantum algorithms and a user-friendly framework. 
Meta-Solving combines three core concepts: the design of \textit{Meta-Solver Strategies}, the reuse and implementation of highly efficient solvers for \textit{Meta-Solver Steps}, and the semi-automatic selection of customizable \textit{Solution Paths}.

\theoremstyle{definition}
\newtheorem{definition}{Definition}[section]

\begin{definition}[Meta-Solver Strategy]
A Meta-Solver Strategy is a decomposition of a mathematical problem into multiple sub-problems, each addressed with tailored algorithms, culminating in a comprehensive solution to the original problem. 
It can describe multiple interchangeable combinations of sub-problems and algorithms, meaning that it can define multiple different methods to solve the problem. 
\end{definition}

\begin{definition}[Meta-Solver Step]
A Meta-Solver Step is a component of a Meta-Solver Strategy, typically associated with a single sub-problem. 
It isolates a specific aspect of a complex mathematical problem for which a dedicated algorithm can be designed and applied to address this segment. 
Each step functions as a building block, contributing a partial solution that, when integrated with others, contributes to a complete solution to the overarching problem.
\end{definition}

\begin{definition}[Solution Path]
A Solution Path represents one specific method to solve a mathematical problem derived from a Meta-Solver Strategy. 
Each Solution Path is a combinations of Meta-Solver Steps that are consecutively executed.
\end{definition}

\begin{figure}[t]
    \centering
    \includegraphics[width = \columnwidth]{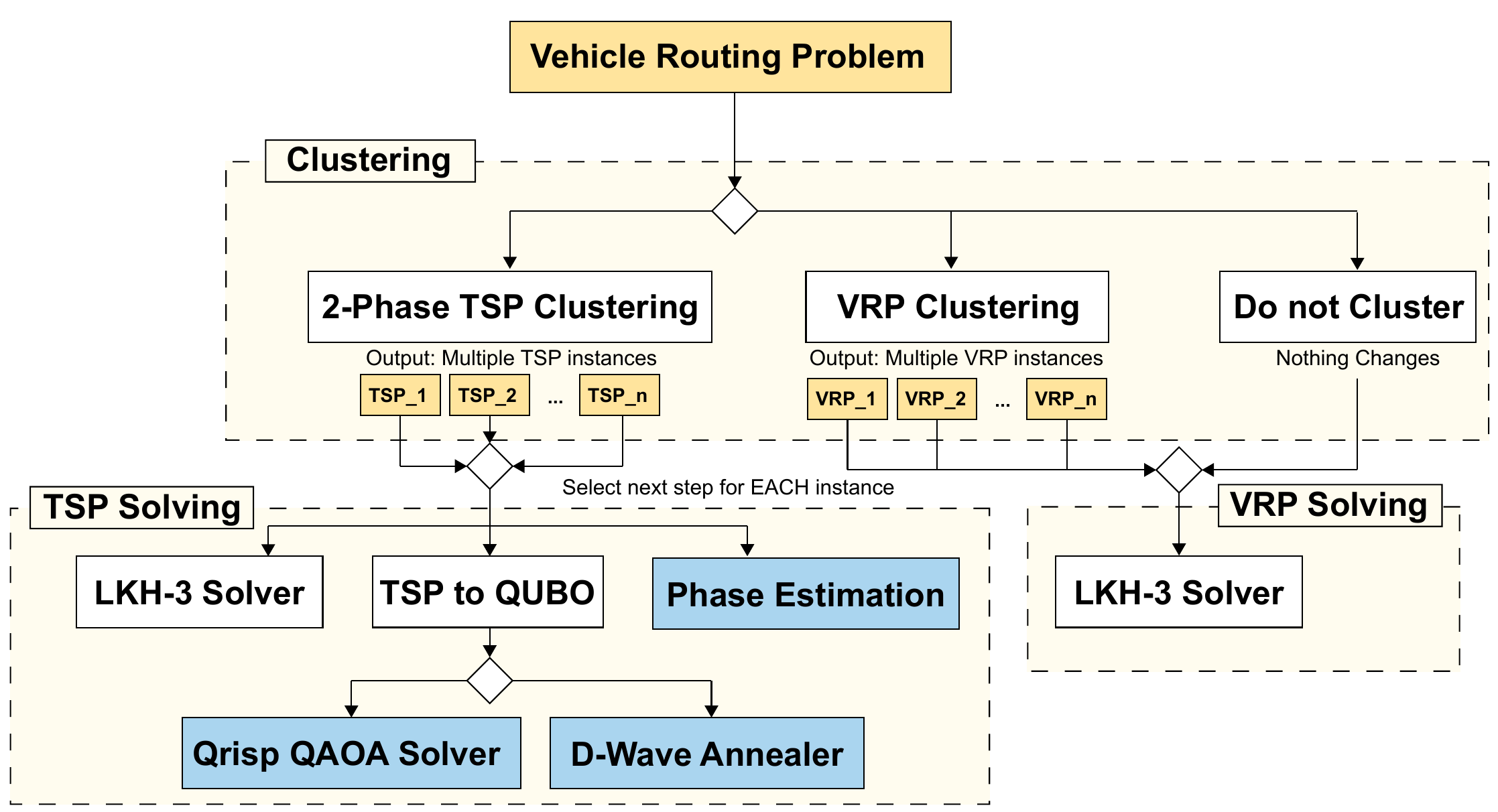}
    \caption{Meta-Solver Strategy for Vehicle Routing Problems.}
    \label{fig:ms_cvrp}
\end{figure}

A Meta-Solver Strategy can be visualized as a tree, where the root node describes the associated mathematical problem, and all other nodes represent the Meta-Solver Steps forming the solving process. 
Nodes that have no children represent the final step of a decomposition, meaning that after executing them, a result for the algorithmic problem can be constructed by recombining the results of the executed steps. 
A Solution Path can be visualized as a path of the tree, starting at the root node, and ending in a leaf. 
The structure of the tree, given by the edges between the nodes, describes which Meta-Solver Steps can and must be combined to find a solution. 
An example of a Meta-Solver Strategy for Vehicle Routing Problems (VRP) is shown in Figure~\ref{fig:ms_cvrp}.
The Meta-Solver Strategy decomposes the problem into two main steps: 
(1) applying a clustering technique, and (2) solving the clusters. 
Depending on the clustering technique used, the clusters are either a set of Traveling Salesperson Problem (TSP) or VRP instances. 
The decomposition shown is inspired by classical state-of-the-art techniques:
VRPs are usually solved by specialized solvers, such as the Lin-Kernighan-Helsgaun (LKH-3) solver~\cite{helsgaun2017extension}. 
If a problem instance is too large to be solved in feasible time, clustering techniques are used to split the problem into several smaller sub-problems, which is faster but may reduce the quality of the solution~\cite{jain1988algorithms}. 

We combine classical and quantum optimization by adding the ability to solve TSP instances by reformulating them as a Quadratic Unconstrained Binary Optimization (QUBO) problem~\cite{lucas2014ising}, which can then be solved with a Quantum Annealer~\cite{2022HybridSF} or QAOA~\cite{ruan2020quantum}.
Another option is solving the TSP instances directly with a phase estimation technique proposed by Srinivasan et. al.~\cite{srinivasan2018efficient}.
Implementations of the QAOA and phase estimation techniques are available in Qrisp~\cite{seidel2022qrisp}, for the Annealing we utilize the D-Wave platform~\cite{d-wave_developers_d-wave_2020}.
All other steps of the Meta-Solver Strategy are classical. 
LKH-3~\cite{helsgaun2017extension} is a solver applicable to TSP and VRP problems. 
For the VRP clustering we use the k-Means method~\cite{jain1988algorithms}. 
LKH-3 and k-Means are well established in classical optimization. 
For the TSP clustering, we choose a specific 2-phase clustering approach by Laporte and Semet~\cite{laporte2002classical}, which has already been studied for hybrid TSP solving~\cite{feld2019hybrid}.
It consists of a creation phase and an improvement phase, and must ensure that each TSP instance can be covered by one truck. 

\begin{figure}[ht]
    \centering
    \includegraphics[width = \columnwidth]{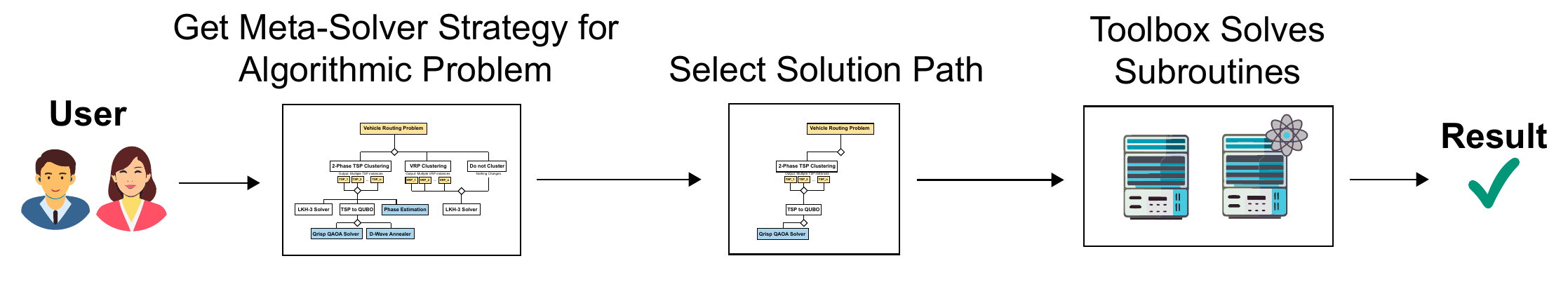}
    \caption{Example process that visualizes how a user applies Meta-Solving.}
    \label{fig:process}
\end{figure}

The concept of Meta-Solving is introduced to the user through the provision of Meta-Solver Strategies, exemplified by the strategy depicted in Figure~\ref{fig:ms_cvrp}. 
In addition, the user is provided with a software tool that implements the strategies and enables the user to engage with them. 
The user should be able to input an algorithmic problem using a standardized format and then semi-automatically select Solution Paths to solve it. 
The software tool can assist the user in selecting a Solution Path, for example by providing the user with expert knowledge about solvers, or by analysing the user's input and suggesting a suitable solver based on established heuristics. We show an example interaction of a user that applies Meta-Solving in Figure~\ref{fig:process}.
Experienced users and researchers may wish to customize a Meta-Solver Strategy or heuristic, or extend a strategy by adding new steps and solvers. 
The following sections explain the core concepts and challenges of a Meta-Solving software tool, and how we envision its application. 

\subsection{Designing and Implementing Meta-Solver Strategies}

The design and implementation of efficient Meta-Solver Strategies is a key necessity to facilitate the access to hybrid solution strategies for end users.
However, finding decompositions that can utilize quantum steps and implementing respective solvers such that they are able to compete with classical state-of-the-art techniques is a challenging task that requires expert knowledge from various fields. 

To create a new Meta-Solver Strategy, an expert must first find a mathematically reasonable decomposition.
A good practice here is to take inspiration from well-established classical solvers.
These are usually well researched and optimized over years, and many of them are open source or free to use for academic purposes. 
They are often benchmarked against each other, and there is knowledge of cases where they perform exceptionally well or bad. 
Our goal is to extend classical optimization techniques with quantum algorithms, so we want to find steps in the classical algorithms that we can exchange with quantum algorithms.
In classical optimization, there are usually many approaches to solving a problem, each with different advantages and disadvantages, and many classical approaches involve several steps where quantum computing could potentially be applied.
Moreover, the possible applications of quantum computing are likely to increase with ongoing research, especially when more modular and open-source optimization frameworks become available in the future. 
With our Meta-Solving approach we want to actively encourage the implementation and comparison of multiple solvers and Meta-Solver Steps. 
We want to promote their unique advantages and enable users to benefit from them. 
Thus, basing Meta-Solver Strategies on the combination of several classical state-of-the-art approaches is a good starting point. 

The implementation of a Meta-Solver Strategy is built from two parts: 1) solvers for the Meta-Solver Steps, and 2) an orchestration unit that can handles the decomposition of the problem.
To implement the solvers, existing highly efficient classical solvers should be used wherever possible.
In our Vehicle Routing Example we reuse the LKH-3~\cite{helsgaun2000effective} solver.
Wrappers will then provide compliance with the necessary input and output formats.
Although the quantum steps are not as advanced, the same concept applies for them. 
Currently, there are frameworks like Qiskit-Optimization~\cite{QiskitOptimization} or Qrisp~\cite{seidel2022qrisp} that provide implementations for well-known quantum algorithms, which should be taken advantage of. 

\begin{figure}
    \centering
    \includegraphics[width = \columnwidth]{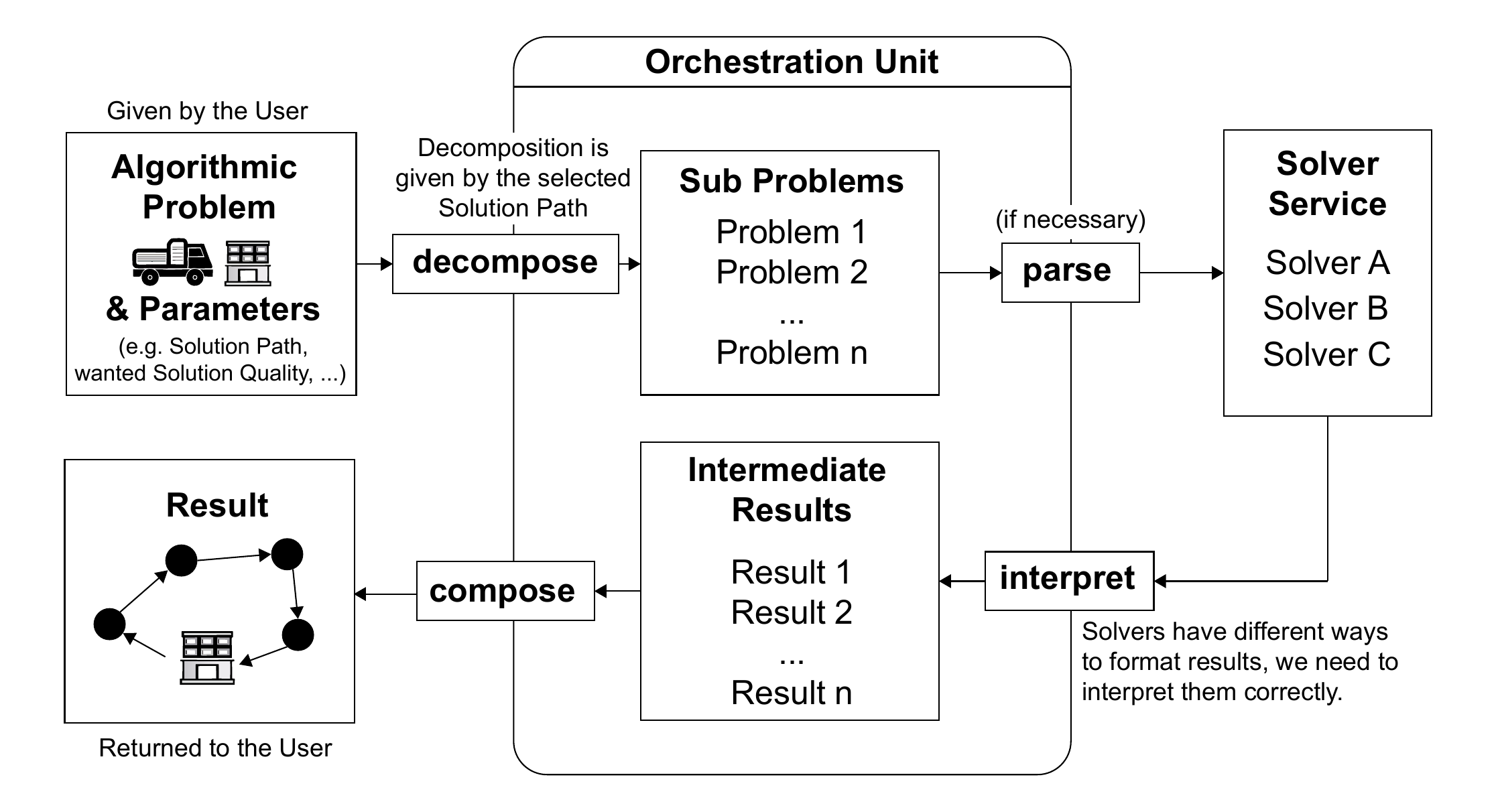}
    \caption{Overview of the Orchestration Unit that decomposes an algorithmic problem to execute a Solution Path, calls the associated Meta-Solver Steps, and then composes the results.}
    \label{fig:orchestration-unit}
\end{figure}

For the orchestration unit, we have to create an interface where an algorithmic problem and additional parameters, such as the Solution Path, can be inserted.
We use standardized formats to define the algorithmic problem, which increases compatibility with existing tools and solvers, allowing for a more modular setup of the pipeline.
An example format for Vehicle Routing Problems is the TSPLIB format~\cite{reinelt1995tsplib95}. 
The orchestration unit decomposes the original problem into multiple sub-problems, each of which is solved by a solver selected in the Solution Path. 
To support solver-specific input formats, additional parsing steps can be inserted.
The orchestration unit must interpret the solution of each solver, which provides a set of intermediate results that are then used to construct a result for the original algorithmic problem. 
Figure~\ref{fig:orchestration-unit} provides an overview of the orchestration unit.

\subsection{Generalized and Specialized Meta-Solving Strategies}

Algorithmic problems exhibit varying degrees of generality, ranging from general mathematical formulations such as Quadratic Unconstrained Binary Optimization (QUBOs) or Integer Linear Programming (ILP) to more specific problems like vehicle routing or the knapsack problem. Similarly, Meta-Solver Strategies can be categorized into general strategies that cover a broad spectrum of problems or specialized strategies tailored for specific problem domains.

Creating a Meta-Solver Strategy necessitates considerable effort, and there might not be a readily available strategy for every problem type. Consequently, having general strategies, such as one designed for Integer Linear Programming, can effectively address a wide array of problems. 
However, specialized solvers can be customized and optimized for a particular problem, enabling performance enhancements tailored to the problem's unique characteristics. 
In such cases, highly specific heuristics may result in superior performance.

For instance, consider the Vehicle Routing Problem again. It can be tackled using a dedicated VRP solver or by reformulating it into a more general problem representation like an ILP, subsequently employing an ILP solver. While the specialized VRP solver is likely to deliver better performance, the ILP solver facilitates easier extension of the problem with additional constraints beyond the scope of VRP definitions. 

In summary, the choice between generalized and specialized Meta-Solver Strategies involves a trade-off between versatility and performance optimization. General strategies are more broadly applicable, but may perform worse than specialized ones.
Consequently, the selection of an appropriate strategy depends on the specific problem requirements and the balance between generality and performance.

\begin{figure}[t]
    \centering
    \includegraphics[width = \columnwidth]{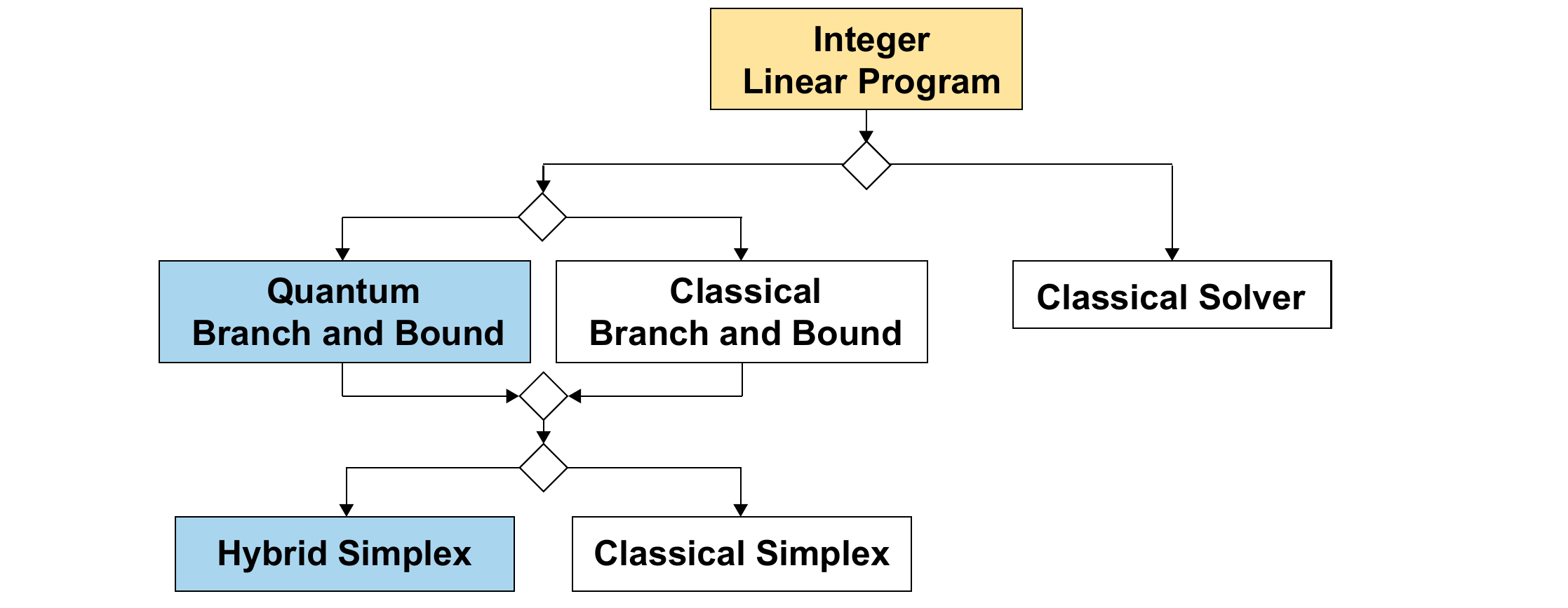}
    \caption{Example Meta-Solver Strategy for Integer Linear Programs.}
    \label{fig:ms_ilp}
\end{figure}

Generalized Meta-Solver Strategies are built similarly to specialized ones. We have shown an example of a specialized strategy in Figure~\ref{fig:ms_cvrp}, which covers VRP solving. An example of a general strategy covering ILP solving is shown in Figure~\ref{fig:ms_ilp}. 
The ILP solution strategy is kept simple, consisting of a classical solution path that does not decompose the problem further, and a hybrid solution path that combines the Branch and Bound~\cite{lawler1966branch} and Simplex~\cite{nabli2009overview} techniques. For both, the application of quantum algorithms has been studied, so an implementation of a quantum branch and bound method~\cite{montanaro2020quantum}, combined with a hybrid Simplex~\cite{nannicini2022fast} is possible and could provide speedups on very advanced quantum hardware. 

\subsection{Input Selection and Interpretation}
\label{sec:GivingAnalyzingInput}

Meta-Solver Strategies offer users the possibility to create a variety of solvers.
However, selecting the most suitable Solution Path entails careful consideration of various factors. 
To assist users in navigating this decision-making process, it is essential that they provide input reflecting their preferences and problem characteristics. 
Users must weigh their priorities, such as solution quality and computational efficiency, when deciding which path to pursue. 
For example, obtaining optimal or near-optimal solutions often requires substantial computational resources, resulting in high costs. However, faster solutions may compromise optimality.

To assist users in navigating this decision-making process, it is essential that they provide input reflecting their preferences and problem characteristics. 
A fundamental input parameter is the algorithmic problem itself. 
By analyzing the problem's syntax, semantics, and characteristics such as size and complexity, we can tailor our approach accordingly. 
Some characteristics, like problem size, readily inform decisions about computational resources, such as whether a problem can be encoded on a quantum computer or whether clustering techniques may be advantageous. 
However, other problem-specific criteria may require a more complex analysis.

As an example, consider the solving of Linear Programs (LPs), which can be solved in many different ways, for instance, by using a Simplex~\cite{nabli2009overview} or Interior-Point~\cite{karmarkar1984new} algorithm. 
The choice between these algorithms depends on factors such as problem size and density. 
For instance, the mentioned Interior-Point algorithm is typically better suited for large, sparsely populated problem instances, whereas the Simplex algorithm works better on small, densely populated instances. 
In this case, the superior algorithm can be inferred from the population characteristics of the LP.

Beyond problem-specific traits, user preferences play an important role.
Providing users with customizable parameters, such as sliders to indicate preferences for speed versus optimality, or information about available hardware resources, enables even more tailored solution recommendations.
However, it is crucial to balance the detail of our analysis with computational efficiency. 
The analysis should be conducted efficiently, ensuring that the time spent evaluating Solution Paths does not overshadow the time required to solve the problem itself.

In summary, facilitating user input and preferences and the analysis of problem characteristics are key components of guiding users towards finding effective Solution Paths. 
By providing users with intuitive interfaces to submit problems and express solution expectations, coupled with efficient analysis techniques, we build a basis that allows us to offer tailored recommendations that optimize the Meta-Solving experience.

\subsection{Suggesting Solution Paths}

Our aim is to empower users to semi-automatically navigate Solution Paths within Meta-Solver Strategies by offering tailored suggestions. 
However, the task of selecting the most optimal solver for a given problem is inherently challenging, even within classical optimization domains. 
The addition of quantum algorithms further complicates this decision-making process. Consequently, we opt to suggest Solution Paths instead of making fully automated selections. 

Given the difficulty of a semi-automated solver selection, we propose to employ diverse heuristics and strategies to facilitate user-guided selection. 
Central to this approach is the input provided by the user, as discussed in sub-section~\ref{sec:GivingAnalyzingInput}. 
Our proposed method involves a step-wise suggestion and selection technique, wherein users are presented with Meta-Solver Strategies and proceed to select Solution Paths incrementally. 
At each step, suggestions from the software toolbox guide the user's decision-making process. Depending on the confidence level of our suggestions, we may present them in various formats. Highly certain suggestions may be presented by clearly highlighting the next recommended step, while less definitive suggestions could be accompanied by a list of pros and cons that the user has to interpret himself.

A significant challenge in suggesting Solution Paths lies in the uncertainty inherent to such recommendations. 
While leveraging heuristics can help in providing informed suggestions, their availability is not guaranteed. 
Consequently, we explore ways for developing new heuristics or employing alternative techniques to assist users in the decision-making process.

One approach it to embedd expert knowledge within Meta-Solver Strategies. 
Each implemented solver would be annotated with detailed information, allowing users to gain insight into the particularities of each solver. 
Access to such distinguished knowledge empowers users to make informed decisions, especially in the absence of readily applicable heuristics.
A disadvantage of this approach is that detailed insights are usually not available for closed-source solvers. 

Additionally, we consider the application of machine learning techniques to derive new heuristics for implemented solvers. 
This involves constructing benchmarking sets that include realistic real-world problem instances and artificially generated challenging instances. 
By evaluating all solvers and potential Solution Paths against these benchmarks, we can train machine learning models which can then be applied to offer recommendations.
Similar to the analysis techniques proposed earlier, we must be careful that the computational cost of machine learning techniques does not outweigh the benefits they provide. 

In conclusion, our approach to facilitating semi-automatic selection of solution paths in Meta-Solver Systems involves a combination of user input, heuristic guidance, expert knowledge embedding, and machine learning techniques. 
By leveraging these strategies, we endeavor to empower users to navigate solution paths effectively within complex optimization domains.

\subsection{Parallel Execution}

In certain scenarios, suggesting Solution Paths within Meta-Solver Strategies may not be feasible. 
This limitation arises when no heuristic is available or when existing heuristics fail to determine a clear preference. 
To address this challenge, various approaches can be employed. 
While randomly selecting a Solution Path is one option, it carries the risk of yielding suboptimal results. 
We propose a more effective strategy that involves executing multiple Solution Paths simultaneously and subsequently comparing their outcomes.

This concept requires Meta-Solving software tools to support parallel execution of multiple Meta-Solver Steps, enabling users with a visualization of multiple results that can be compared.
Now, a new challenge emerges: determining the best solution among those presented. 
Although implementing assessment techniques to grade solution quality is possible, it remains a challenge to determine a result that best fits the users needs in a general case. 
Thus, we again refrain from automating this step and instead enable users to make their own informed decisions. 
We assist users in this process by providing comprehensive visualizations and combine them with assessment metrics.
Furthermore, we have to consider that some solvers, particularly quantum ones, are non-deterministic. 
Consequently, they may produce different solutions when executed multiple times. 
To accommodate this variability, our approach incorporates the option to apply multiple trials for both quantum and classical solvers.

It is important to state that the parallel execution and comparison technique entails a notable drawback: it significantly increases computational effort. 
Users must therefore make informed decisions and carefully consider scenarios where parallel execution and result comparison are warranted.

The combination of parallel execution and heuristic guidance offers a promising framework for identifying optimal Solution Paths. 
This approach not only facilitates the creation of highly efficient solvers but also enables the utilization of quantum algorithms, thereby enhancing the Meta-Solver Strategies' effectiveness.

\subsection{Backend Selection}

Upon selecting a Solution Path within Meta-Solver Strategies, the subsequent execution of included Meta-Solver Steps necessitates compatibility with appropriate backends. 
For classical algorithms, these backends typically encompass GPU or CPU computing clusters, whereas for quantum algorithms, options include simulators, quantum annealers, or universal quantum computers of various hardware technologies. 
Notably, while quantum simulators serve well for testing and learning purposes, they  fall short in achieving actual speedups.

The choice of backend holds significant implications for computation efficiency and solution quality. 
To assist users in this process, a semi-automated backend selection should be supported within Meta-Solving.
An application of such an approach first requires users to provide information about the backends they have access to.
Accessing classical computation backends is generally straightforward, as individuals can book computation time, and institutions such as companies or universities typically possess readily available resources.
However, access to quantum computers remains more limited, 

For classical algorithms, determining the essential computational resources for a solver (e.g., memory, GPU/CPU power, cores) is also straightforward. 
This information is typically gained when executing the solver on sample sets. 
By annotating this information for each implemented classical solver, we can propose a fitting classical computation cluster accordingly.

In contrast, quantum backend selection is notably more intricate due to the distinctive characteristics of quantum hardware. 
Factors such as varying technologies for qubit representation, programming methods, unique qubit mappings, error rates, and optimization requirements for specific low-level hardware pose significant challenges. Exact error rates even vary over time due to calibration uncertainty.
Consequently, selecting a well-suited quantum backend assumes paramount importance.

This complexity is addressed by sophisticated quantum backend selection frameworks such as the MQT Predictor~\cite{quetschlich2023mqt} or the  NISQ analyzer~\cite{salm2020nisq}, which can be integrated into the orchestration units of a Meta-Solving software framework.
By leveraging prior work in this field, we can develop robust algorithms and frameworks to guide users in selecting appropriate quantum backends, thereby enhancing the efficacy of Meta-Solver Strategies in quantum and classical computing domains.

\subsection{Composing Meta-Solver Strategies}

Algorithmic problems and Meta-Solver Strategies operate at different levels of abstraction and generality. 
For instance, sorting problems represent low-level, general challenges, whereas Vehicle Routing embodies a specialized, high-level problem. 
Low-level algorithms often solve sub-problems of high-level problems, prompting the reuse of low-level Meta-Solver Strategies within higher-level ones. 
This technique, which we call Composing Meta-Solver Strategies, facilitates the integration of existing strategies as Meta-Solver Steps in more complex strategies.

As an example, consider a Meta-Solver Strategy designed to solve QUBO problems.
This strategy can be repurposed to tackle vehicle routing problems and integer linear programs. 
For instance, in the VRP Meta-Solver Strategy from Figure~\ref{fig:ms_cvrp}, instead of defining a customized QUBO solving step, we can simply call the existing QUBO strategy and continue the solution process. 
Similarly, in the ILP strategy from Figure~\ref{fig:ms_ilp}, which currently lacks a QUBO solving step, we could extend it by adding a step to convert the ILP into a QUBO~\cite{chang2020integer}.

Composing Meta-Solver Strategies offers substantial time-saving benefits by leveraging existing implementations and heuristics. 
However, there are scenarios where reusing a strategy may not be advantageous. 
For example, one approach to solving QUBOs involves the Variational Quantum Eigensolver (VQE)~\cite{peruzzo2014variational}, whose performance depends heavily on finding a suitable ansatz, which can be highly problem specific. 
Implementing a general-purpose VQE and reusing it across different QUBOs can lead to performance degradation.
In this example, a user needs to be guided in two decisions. 
First, he must decide whether reformulating the problem as a QUBO is beneficial to the solution process, and if so, he must decide which QUBO solving method is best to solve his specific problem. 

In conclusion, it is crucial to strike a balance between when reusing general implementations proves beneficial and when it is detrimental. 
By carefully evaluating the specific characteristics of the problem domain and the performance implications of reuse, practitioners can effectively leverage Capsuling of Meta-Solver Strategies to optimize solution processes.



\section{Implementation}
\label{sec:implementation}

In this section, we present how our ProvideQ toolbox prototype \cite{eichhorn2023providing} implements the concepts set out in Section~\ref{sec:concept1}.
This implementation allows us to evaluate the concept in Section~\ref{sec:evaluation} and enables users to explore the Meta-Solver Strategy.

The ProvideQ toolbox is structured with a client-server architecture where the server is responsible for data retention and solver execution and the client serves as a user interface. 
This separation, in combination with a well-documented REST API, also ensures that the toolbox can be controlled automatically and by other applications.
The data structures and processes within the ProvideQ toolbox closely resemble the key definitions presented in Section~\ref{sec:concept1}.
Both the root problem and every step of a Meta-Solver Strategy are represented as \texttt{Problem}s, a tree-like recursive data structure.
Starting with an empty root problem node, the toolbox automatically expands the problem tree with problem branches corresponding to the tasks required for executing a selected solver.
This recursive data structure enables the toolbox to solve steps independently and in parallel.

The design of the toolbox is centered around the definition and implementation of problem solvers.
Problem solvers need to implement a simple interface which provides them with input data and requires them to return output data.
Meta-Solver Strategies and orchestration units are implemented as separate layers on top of the problem solver layer.
This separation ensures that developers implementing problem solvers do not need to know about the details of Meta-Solver Strategies.
Furthermore, the toolbox library provides various utility modules assisting problem solver implementations to integrate external tools and languages.
For example, the ProvideQ toolbox supports the usage of the Qiskit and Qrisp frameworks by providing a \texttt{PythonProcessRunner} for running Python scripts, writing input files, and reading output files.
These frameworks, in turn, can be used to execute code on external hardware, like running quantum circuits on quantum computers.
Additionally, the toolbox server can read and write problem instances in well-known, standardized formats like the TSPLIB format for Vehicle Routing Problems~\cite{reinelt1995tsplib95}, or the 
DIMACS format for Boolean Satisfiability (SAT) problems~\cite{johnson1996cliques}.
This ensures the toolbox can be easily integrated with existing solvers, both when using the toolbox in an external application and when integrating existing solvers into to toolbox. 

The toolbox client presents the features of the server in a web application.
This component utilizes the server's API to visualize Meta-Solver Strategies, enabling users to submit problems, interactively select Solution Paths, and compute solutions.
The user is guided through the problem-solving process with Solution Path suggestions.

\begin{figure}
    \centering
    \includegraphics[width = \columnwidth]{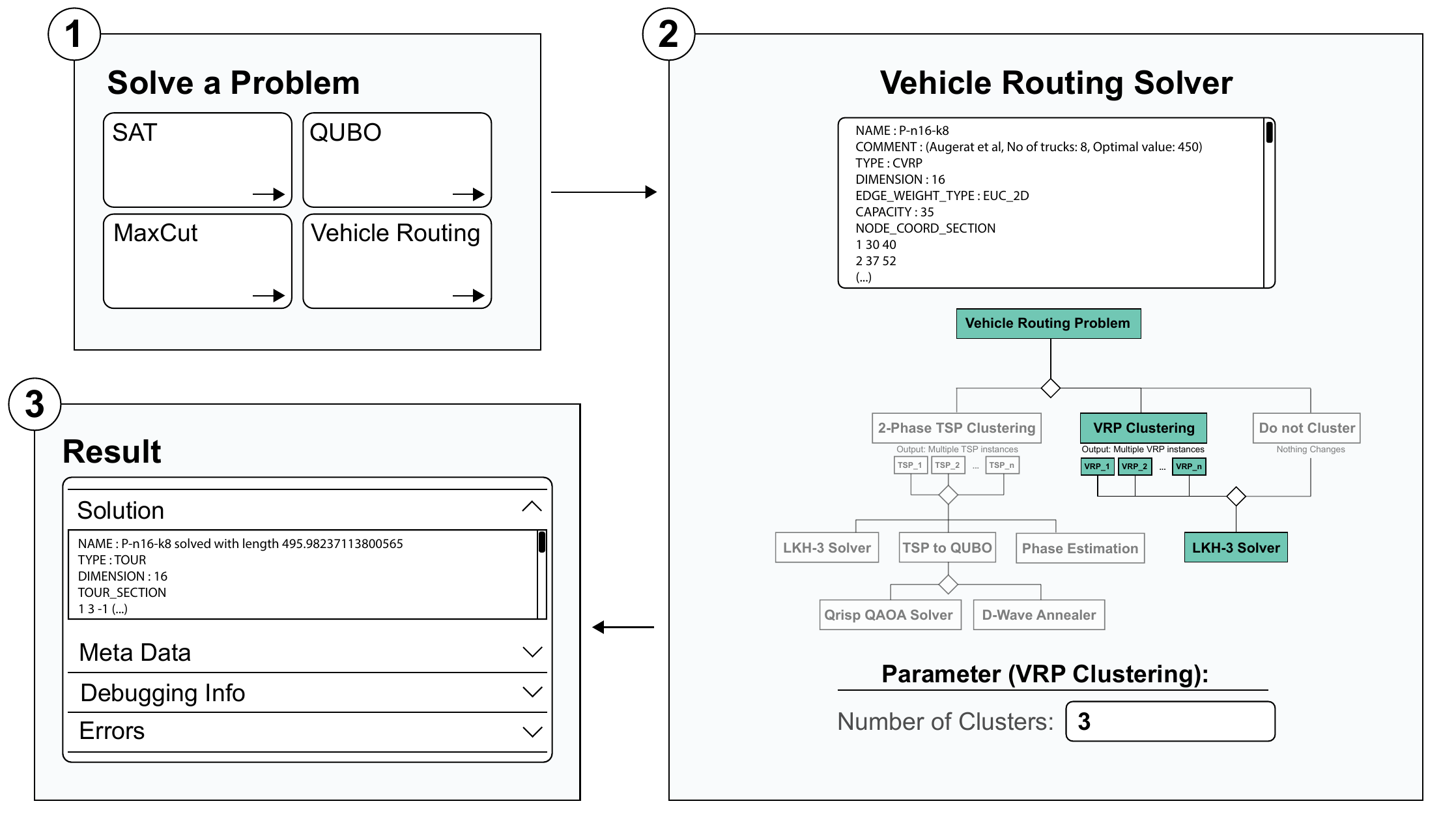}
    \caption{User Interface mockup of the ProvideQ Toolbox. It combines the Vehicle Routing example from Figure~\ref{fig:ms_cvrp} with the workflow from Figure~\ref{fig:process}}
    \label{fig:prototype}
\end{figure}

The problem-solving workflow involves three steps in the ProvideQ web interface, as shown in Figure~\ref{fig:prototype}.
First, the user chooses a specific problem to solve from the list of available problems.
Currently available problems in the prototype include VRP, QUBO, SAT and MaxCut.
Second, the toolbox prompts the user to enter the problem instance in a standardized text format into the input field.
Additionally, the Meta-Solver Strategy of the selected problem is visualized as a tree.
Here, each node represents a Meta-Solver Step, and the user needs to select solvers consecutively to build a Solution Path for the Meta-Solver Strategy.
Third, the user can inspect the current state, results, and solver specific settings of each step in separate views.
In some cases, Meta-Solver Steps require specific settings, for example, the desired number of clusters in a VRP clustering step.
The user can choose between a stepwise or complete execution of the Solution Path.
When partially solving a Solution Path, inspecting specific Meta-Solver Step results can guide the decision process to select the next step.
For example, the current clustering of a VRP problem might still be too large to be applicable for a Quantum solver and based on this insight another clustering step is inserted into the Solution Path.



At this point in time, the ProvideQ toolbox is a prototype for Hybrid Meta-Solving.
It is developed as an open-source project on GitHub\footnote{GitHub repository: \url{https://github.com/ProvideQ}} and currently implements problem definitions, Meta-Solver decomposition, Composing of strategies, and the orchestration unit.
Additionally, the interactive tree-like user interface for Solution Paths is currently in development.
The open-source model of the toolbox allows users to incorporate new problems and solvers and share them with others at their own discretion.
This way, experienced users can host their own instances of the toolbox-server to leverage their computational resources and to customize the toolbox depending on their individual requirements.
Many more ideas are yet to be implemented, for example the semi-automated Solution Path and backend suggestions, and the sophisticated parallel execution of Solution Paths, as described in Section~\ref{sec:concept1}.
As the ProvideQ toolbox is designed with these extensions in mind, we leave these tasks to be implemented in future work.


\section{Evaluation}
\label{sec:evaluation}

We introduced a new technique for solving optimization problems: Hybrid Meta-Solving. 
We want to evaluate if our technique is implementable and if it is able to fully exploit state-of-the-art classical techniques.
Furthermore, we want to study how the introduction of quantum steps in a Meta-Solver Strategy can affect the solving process and the obtained results. 
The focus of the evaluation is not to outperform any classical state-of-the-art, as this is not achievable with our currently available quantum simulations and hardware.
Rather, we want to prove that our concept can reuse, extend, and combine classical techniques with quantum computing methods.
More specifically, we want to answer the following research questions (RQ):

\begin{itemize}
    \item \textbf{RQ1:} How can classical state-of-the-art techniques be adapted in a Meta-Solver Strategy?
    \item \textbf{RQ2:} How do Solution Paths with quantum subroutines perform compared to purely classical Solution Paths?
\end{itemize}

\subsection{Experiment Design}

To perform the evaluation, the Vehicle Routing Meta-Solver Strategy, as depicted in Figure~\ref{fig:ms_cvrp}, was implemented and integrated into the ProvideQ toolbox. 
The evaluation was performed on 23 capacitated Vehicle Routing Problems provided through the CVRPLIB website~\cite{uchoa2017new}, for which the optimal solution is already known.
We utilized Set-P, provided by Augerat et al.~\cite{augerat1995computational}, due to its inclusion of both smaller (fewer than 25 cities) and larger instances (with more than 70 cities). 
The names of the problems indicate their major characteristics. The number of cities is indicated by n, while the number of available trucks is indicated by k. For instance, p-n50-k7 is a problem with 50 cities (one of which is the depot) and 7 trucks. 
The evaluation was conducted on a MacBook Pro from 2021 with an M1 chip and 32 gigabytes of RAM. 
The quantum steps were simulated, rather than executed on a quantum computer. 

Assuming that we only cluster once, we could create a total of six Solution Paths, four of which will be examined further:

\begin{itemize}
    \item [(1)] No clustering + LKH-3 Solver, 
    \item [(2)] 2-Phase TSP clustering + LKH-3 Solver, 
    \item [(3)] 2-Phase TSP clustering + Qrisp QAOA Solver, 
    \item [(4)] 2-Phase TSP clustering + D-Wave Annealer.
\end{itemize}

Solution Paths (1) and (2) are purely classical, whereas Solution Paths (3) and (4) combine a classical clustering technique with hybrid QUBO solving techniques.
We did not consider the VRP clustering in combination with the LKH-3 Solver, as for the selected benchmarking instances, a clustering was not necessary to successfully apply LKH-3. 
However, we did combine LKH-3 with the 2-Phase TSP clustering to enable a more insightful comparison between purely classical and hybrid solvers. This was necessary because the 2-Phase TSP clustering had to be applied to execute the QAOA and Annealing Solver.
We did not consider the Phase Estimation Solver because it could only be applied to TSP instances containing a maximum of five cities.
Even after applying the 2-Phase TSP clustering, the majority of derived clusters were larger than five cities, preventing the identification of solutions for all problems instances expect two.
We faced a similar issue with the QAOA solver, and therefore decided only to include one of them in the data shown below because those results are not meaningful in the presented context.
We also tried to address this issue by applying the clustering multiple times, but this resulted in highly inefficient solutions.

\begin{figure}[ht]
    \centering
    \includegraphics[width = \columnwidth]{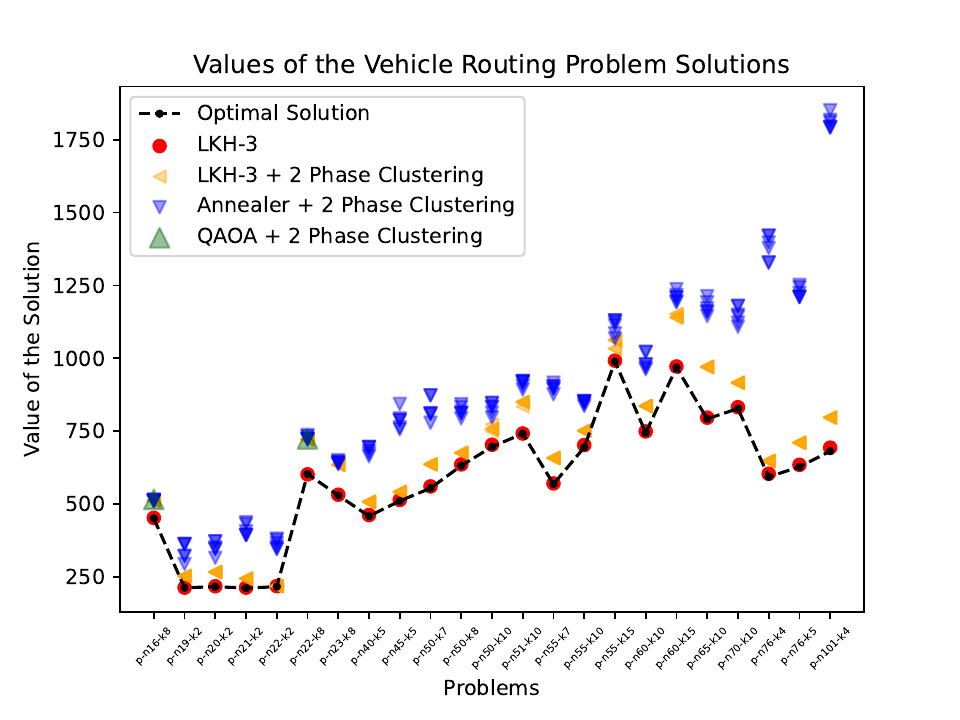}
    \caption{The plot illustrates the cost of the solution retrieved for all Solution Paths. 
    The cost of the solution is calculated by adding up the total number of kilometers to be traveled, with lower values representing a more optimal solution. The scattered black line represents the optimal solution of the problem. The plot demonstrates that the LKH-3 solver (without clustering) was able to compute nearly optimal solutions. Conversely, Solution Paths with clustering yielded worse results. Notably, the QAOA path only produced solutions for two problems.}
    \label{fig:data_solution_quality}
\end{figure}

\begin{figure}[t]
    \centering
    \includegraphics[width = \columnwidth]{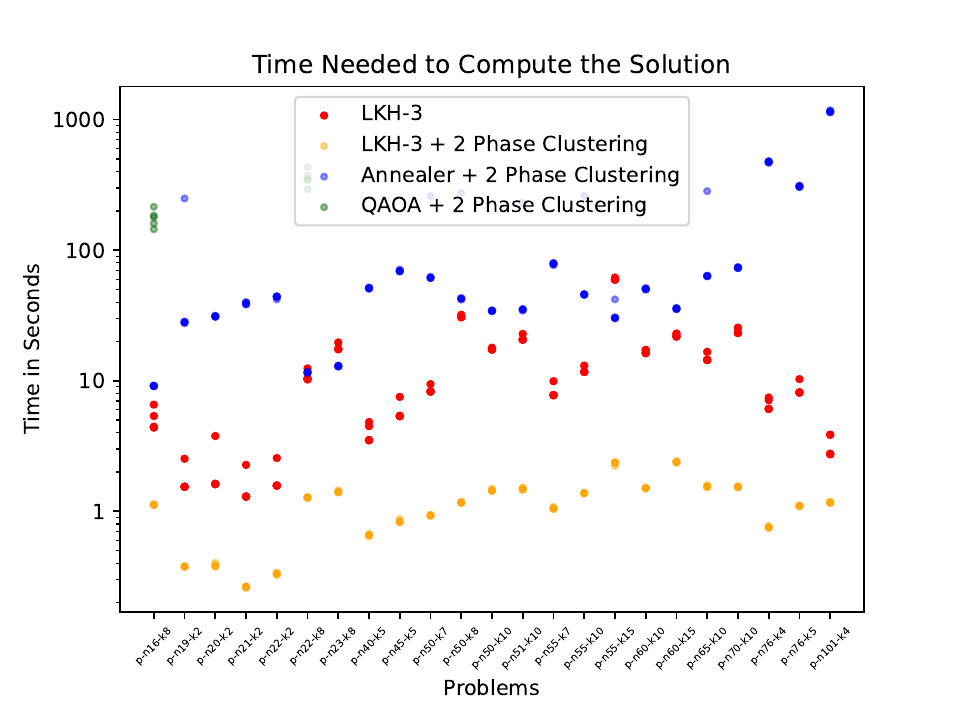}
    \caption{A logarithmic plot of the time required to solve a problem. 
    It reveals that the addition of a clustering step can significantly reduce the calculation time for the LKH-3 solver. In contrast, the calculation times of the Annealer and QAOA paths are considerably higher, largely due to the quantum simulation overhead included here.
    A direct comparison between a classical solver and a simulated quantum solver is not a meaningful approach for evaluating their relative effectiveness. However, we have included those values for transparency.}
    \label{fig:data_runtime}
\end{figure}

All solution paths are executed five times to account for the varying computation times and non-deterministic steps of the algorithms. 
We compare them by measuring their solution quality and execution time. The results are shown in Figure~\ref{fig:data_solution_quality} and Figure~\ref{fig:data_runtime}. 

\subsection{Results}

Answering RQ1, we demonstrated that existing monolithic classical solvers, such as LKH-3, can be repurposed for use in Meta-Solver Strategies. When employed as a standalone solver, LKH-3 yields highly satisfactory results and produces nearly optimal solutions in a matter of seconds for each problem instance.
Furthermore, the Meta-Solver Strategy permitted the integration of the LKH-3 solver with a clustering approach. As anticipated, the incorporation of a clustering step typically results in a reduction in solution quality, yet it simultaneously leads to a significant reduction in computation times. These time savings may become even more pronounced in the context of more complex problem instances.

It is evident that the standalone LKH-3 and LKH-3 with clustering offer distinct advantages when addressing the problem at hand. Standalone LKH-3 necessitates a greater investment of time, yet it is associated with the generation of superior solutions. In contrast, LKH-3 with clustering requires less time, yet it is associated with the generation of inferior solutions. 
By communicating this information to the user, it becomes possible for them to make an informed decision regarding the approach that best suits their needs. 

Answering RQ2, our Meta-Solving approach allows for the straightforward exchange of solvers and integration of quantum computing techniques. 
The 2-Phase TSP clustering was necessary for the retrieval of results, as the original problems are too large to be solved in a simulated environment.
A direct comparison between LKH-3 and the quantum subroutines is not entirely fair, therefore we do not compare the standalone LKH-3 with any of the hybrid paths.

Interesting observations can be made when comparing the LKH-3 with clustering and the annealer with clustering paths. 
Both apply the same clustering technique and start with the same input. 
While the simulated annealer always performs worse than LKH-3, there are some cases where its solution is very close. 
Even though the hybrid approach entails a significant amount of simulation overhead, it has been demonstrated that competitive solution can be retrieved in certain instances.
However, the annealing approach has also exhibited instances where it performed considerably worse, particularly for larger problem instances, especially for the problems with more than 70 instances.
The QAOA technique has an even greater simulation overhead than the annealing method, and therefore only yielded solutions for two problems.
Both problems exhibited a favorable ratio between cities and trucks, with a greater number of trucks available than in other problems. 
Consequently, the 2-Phase clustering technique resulted in the formation of smaller clusters, which enabled the simulation to proceed. 
The results obtained by the QAOA approach were comparable to those obtained by LKH-3 and the Annealer.

It can be concluded that Solution Paths containing quantum subroutines performed worse than purely classical solution paths, even when the same clustering method was applied to them. 
This result was anticipated and is largely attributed to the overhead introduced to enable the application of the quantum simulations. 
There is still a considerable effort needed before quantum methods can be considered competitive with their classical counterparts. 
Nevertheless, our meta-solving frameworks allow for the immediate utilisation of such methods once they become available.


\section{Discussion and Limitations}
\label{sec:discussion}

\subsection{Limitations of Quantum Computing}
Meta-Solver Strategies that involve quantum computing steps are only as powerful as existing quantum algorithms and hardware allow. It is reasonable to imagine that quantum computers can outperform classical computers in certain tasks, and therefore reasonable to claim that we can exploit these advantages with our approach.
However, in the current NISQ era, we are still far from achieving quantum supremacy in optimization. 
As a result, it may take years or even decades before the potential of quantum steps can truly shine. 
However, our Meta-Solvers are built on state-of-the-art classical techniques that can be used without quantum hardware. 
Thus, we built a foundational platform that will be ready for the future advances that quantum computing can bring. 

\subsection{Orchestrational Overhead}

A key part of our Meta-Solving approach is the solution of multiple Meta-Solver Steps, each of which is executed by different types of solvers, and in many cases on different hardware, such as a quantum computer. 
Allowing highly customized solution path choices can have the disadvantage of over-decomposing and re-composing problems, leading to additional orchestration overhead. 
Monolithic state-of-the-art solvers require less orchestration and therefore save some computational overhead. 
However, even with this orchestration overhead, meaning parsing files into different formats or sending a solving request to a compute cluster, is usually only a small overhead. 
We argue that this overhead is negligible because it usually takes only a few seconds or even milliseconds to transform and send this kind of data. 
However, there is one exception to this argument, and that is the transformation of classical data into quantum circuits. 
Even if there is a problem formulation that is advantageous for a quantum computer, such as a parametrized circuit that represents an Ising model, there may still be significant overhead in finding an embedding for the quantum circuit on the actual hardware. 
These kinds of problems are often seen in quantum computing and are a problem that comes with the limitations of current NISQ hardware. 
We believe that embedding problems on a quantum computer will be much more efficient once scalable hardware, error mitigation, and better abstraction layers and compilers are available, so this problem should be solved in the future. 

\subsection{Implementation of the Platform}

The design and implementation of Meta-Solver Strategies is a challenging task. 
First, we have to find mathematically reasonable decompositions, and then we have to implement highly efficient solvers (quantum and classical), analysis techniques for recommending Solution Paths, and user interface/user experience features to allow others to easily interact with our platform. 
We have already implemented a prototype of a Meta-Solving platform and shown that the concepts we have presented are feasible.
However, creating an industry-ready platform that includes a wide variety of Meta-Solver Strategies still requires more work. 

\section{Conclusion}
\label{sec:conclusion}
This paper introduces a novel technique, Hybrid Meta-Solving, which fuses the strengths of classical and quantum optimization. It decomposes mathematical problems into multiple sub-problems and implements a software framework that enables the seamless exchange and extension of solvers for the sub-problems. 
The core concepts of Meta-Solving were introduced, including the semi-automated suggestion of Solution Paths based on user input and problem characteristics, the parallel execution and comparison of Meta-Solver Steps, and an automated backend selection. Additionally, the design and implementation of Meta-Solver Strategies were described, along with their varying degrees of generality and the improvement in reusability that capsuling strategies provide. 
The majority of our concepts were implemented in the ProvideQ toolbox prototype, which allows users to engage with Meta-Solver Strategies and select Solution Paths in interactive workflows. 
Our evaluation demonstrated that Meta-Solver Strategies can reuse existing state-of-the-art solvers and leverage excellent results for the Vehicle Routing examples. Furthermore, it allows us to configure different Solution Paths that leverage different advantages, such as high solution quality or rapid solution generation.
We demonstrated that the incorporation of quantum algorithms is feasible, providing an accessible way to utilize quantum computing techniques.
However, we also observed that Solution Paths that include quantum steps yield inferior results compared to purely classical Solution Paths, rendering the application of quantum computing impractical for these examples. 
We hope that further advances in the field of quantum computing will result in a change of these results, rendering quantum computing a more competitive or even superior alternative to classical state-of-the-art solvers. 
Our framework is capable of providing a fundamental platform for hybrid optimization and will become competitive once future advances in quantum computing are made. 

In future work, we intend to delve more deeply into the subjects of semi-automated Solution Path suggestions, the parallel execution of Meta-Solver Steps, and automated backend selection and circuit optimization. This paper presented the fundamental concepts underlying these techniques, but there is a great deal of research that is necessary to implement them in a general context. 

\section*{Acknowledgment}

The authors would like to thank Ina Schaefer for her valuable input to discussions, and Lucas Berger for helping to implement the presented evaluation.
This work has been supported by the German Federal Ministry for Economics and Climate Action (BMWK) in the projects ProvideQ (reference numbers: 01MQ22006D and 01MQ22006F) and QuaST (reference number: 01MQ22004D). 

We acknowledge that we used AI tools to improve the grammar of this paper. 
We used DeepL Write for basic text editing and occasionally prompted ChatGPT to help with repharsing sentences. Some icons in our Figures were created with Adobe Firefly.  

\balance

\bibliographystyle{IEEEtran}
\bibliography{bibliography}

\begin{thebibliography}{10}
\providecommand{\url}[1]{#1}
\csname url@samestyle\endcsname
\providecommand{\newblock}{\relax}
\providecommand{\bibinfo}[2]{#2}
\providecommand{\BIBentrySTDinterwordspacing}{\spaceskip=0pt\relax}
\providecommand{\BIBentryALTinterwordstretchfactor}{4}
\providecommand{\BIBentryALTinterwordspacing}{\spaceskip=\fontdimen2\font plus
\BIBentryALTinterwordstretchfactor\fontdimen3\font minus \fontdimen4\font\relax}
\providecommand{\BIBforeignlanguage}[2]{{%
\expandafter\ifx\csname l@#1\endcsname\relax
\typeout{** WARNING: IEEEtran.bst: No hyphenation pattern has been}%
\typeout{** loaded for the language `#1'. Using the pattern for}%
\typeout{** the default language instead.}%
\else
\language=\csname l@#1\endcsname
\fi
#2}}
\providecommand{\BIBdecl}{\relax}
\BIBdecl

\bibitem{grover1996fast}
L.~K. Grover, ``A fast quantum mechanical algorithm for database search,'' in \emph{Proceedings of the twenty-eighth annual ACM symposium on Theory of computing}, 1996, pp. 212--219.

\bibitem{shor1999polynomial}
P.~W. Shor, ``Polynomial-time algorithms for prime factorization and discrete logarithms on a quantum computer,'' \emph{SIAM review}, vol.~41, no.~2, pp. 303--332, 1999.

\bibitem{deutsch1992rapid}
D.~Deutsch and R.~Jozsa, ``Rapid solution of problems by quantum computation,'' \emph{Proceedings of the Royal Society of London. Series A: Mathematical and Physical Sciences}, vol. 439, no. 1907, pp. 553--558, 1992.

\bibitem{stilck2021limitations}
D.~Stilck~Fran{\c{c}}a and R.~Garcia-Patron, ``Limitations of optimization algorithms on noisy quantum devices,'' \emph{Nature Physics}, vol.~17, no.~11, pp. 1221--1227, 2021.

\bibitem{bharti2022noisy}
K.~Bharti, A.~Cervera-Lierta, T.~H. Kyaw, T.~Haug, S.~Alperin-Lea, A.~Anand, M.~Degroote, H.~Heimonen, J.~S. Kottmann, T.~Menke \emph{et~al.}, ``Noisy intermediate-scale quantum algorithms,'' \emph{Reviews of Modern Physics}, vol.~94, no.~1, p. 015004, 2022.

\bibitem{farhi2014quantum}
E.~Farhi, J.~Goldstone, and S.~Gutmann, ``A quantum approximate optimization algorithm,'' \emph{arXiv preprint arXiv:1411.4028}, 2014.

\bibitem{osaba2024hybrid}
E.~Osaba, E.~Villar-Rodriguez, A.~Gomez-Tejedor, and I.~Oregi, ``Hybrid quantum solvers in production: how to succeed in the nisq era?'' \emph{arXiv preprint arXiv:2401.10302}, 2024.

\bibitem{Qiskit}
{Qiskit contributors}, ``Qiskit: An open-source framework for quantum computing,'' 2023.

\bibitem{bergholm2022pennylane}
V.~Bergholm \emph{et~al.}, ``Pennylane: Automatic differentiation of hybrid quantum-classical computations,'' 2022.

\bibitem{seidel2022qrisp}
R.~Seidel, S.~Bock, N.~Tcholtchev, and M.~Hauswirth, ``Qrisp: A framework for compilable high-level programming of gate-based quantum computers,'' \emph{PlanQC-Programming Languages for Quantum Computing}, 2022.

\bibitem{eichhorn2023providing}
D.~Eichhorn, M.~Schweikart, N.~Poser, T.~Osborne, and I.~Schaefer, ``Providing quantum readiness: The vision of the provideq toolbox,'' in \emph{INFORMATIK 2023 - Designing Futures: Zukünfte gestalten}.\hskip 1em plus 0.5em minus 0.4em\relax Bonn: Gesellschaft für Informatik e.V., 2023, pp. 1129--1133.

\bibitem{preskill_quantum_2018}
J.~Preskill, ``Quantum {Computing} in the {NISQ} era and beyond,'' \emph{Quantum}, vol.~2, p.~79, Aug. 2018.

\bibitem{abughanem2023nisq}
M.~AbuGhanem and H.~Eleuch, ``Nisq computers: A path to quantum supremacy,'' \emph{arXiv preprint arXiv:2310.01431}, 2023.

\bibitem{peruzzo2014variational}
A.~Peruzzo, J.~McClean, P.~Shadbolt, M.-H. Yung, X.-Q. Zhou, P.~J. Love, A.~Aspuru-Guzik, and J.~L. O’brien, ``A variational eigenvalue solver on a photonic quantum processor,'' \emph{Nature communications}, vol.~5, no.~1, p. 4213, 2014.

\bibitem{poggel_recommending_2023}
\BIBentryALTinterwordspacing
B.~Poggel, N.~Quetschlich, L.~Burgholzer, R.~Wille, and J.~M. Lorenz, ``Recommending {Solution} {Paths} for {Solving} {Optimization} {Problems} with {Quantum} {Computing},'' in \emph{2023 {IEEE} {International} {Conference} on {Quantum} {Software} ({QSW})}.\hskip 1em plus 0.5em minus 0.4em\relax IEEE, Jul. 2023. [Online]. Available: \url{http://dx.doi.org/10.1109/QSW59989.2023.00017}
\BIBentrySTDinterwordspacing

\bibitem{palackal_quantum-assisted_2023}
L.~Palackal, B.~Poggel, M.~Wulff, H.~Ehm, J.~M. Lorenz, and C.~B. Mendl, ``Quantum-{Assisted} {Solution} {Paths} for the {Capacitated} {Vehicle} {Routing} {Problem},'' 2023, \_eprint: 2304.09629.

\bibitem{quantagonia_hybridsolver}
\BIBentryALTinterwordspacing
``\BIBforeignlanguage{en}{{HybridSolver} for advanced {MIPs}, {LPs}, and {QUBOs}: {Quantagonia}}.'' [Online]. Available: \url{https://www.quantagonia.com/hybridsolver}
\BIBentrySTDinterwordspacing

\bibitem{d-wave_developers_d-wave_2020}
D.-W. Developers, ``D-{Wave} {Hybrid} {Solver} {Service}: {An} {Overview},'' D-Wave Systems Inc., Tech. Rep. 14-1039A-B, 2020.

\bibitem{planqk}
\BIBentryALTinterwordspacing
``\BIBforeignlanguage{en}{{PlanQK} {Platform}}.'' [Online]. Available: \url{https://platform.planqk.de/}
\BIBentrySTDinterwordspacing

\bibitem{klau_autoqml_2023}
\BIBentryALTinterwordspacing
D.~Klau, H.~Krause, D.~Kreplin, M.~Roth, C.~K. Tutschku, and M.-A. Zöller, ``{AutoQML} - a {Framework} for {Automated} {Quantum} {Machine} {Learning},'' 2023. [Online]. Available: \url{https://publica.fraunhofer.de/handle/publica/458985}
\BIBentrySTDinterwordspacing

\bibitem{concorde}
\BIBentryALTinterwordspacing
W.~Cook \emph{et~al.}, ``Concorde tsp solver.'' [Online]. Available: \url{https://www.math.uwaterloo.ca/tsp/concorde/index.html}
\BIBentrySTDinterwordspacing

\bibitem{Koch2022}
\BIBentryALTinterwordspacing
T.~Koch, T.~Berthold, J.~Pedersen, and C.~Vanaret, ``Progress in mathematical programming solvers from 2001 to 2020,'' \emph{EURO Journal on Computational Optimization}, vol.~10, p. 100031, 2022. [Online]. Available: \url{http://dx.doi.org/10.1016/j.ejco.2022.100031}
\BIBentrySTDinterwordspacing

\bibitem{bussieck2011}
\BIBentryALTinterwordspacing
M.~R. Bussieck and S.~Vigerske, \emph{MINLP Solver Software}.\hskip 1em plus 0.5em minus 0.4em\relax John Wiley \& Sons, Ltd, 2011. [Online]. Available: \url{https://onlinelibrary.wiley.com/doi/abs/10.1002/9780470400531.eorms0527}
\BIBentrySTDinterwordspacing

\bibitem{Kallrath2011}
\BIBentryALTinterwordspacing
J.~Kallrath, ``Polylithic modeling and solution approaches using algebraic modeling systems,'' \emph{Optimization Letters}, vol.~5, no.~3, pp. 453--466, Aug 2011. [Online]. Available: \url{https://doi.org/10.1007/s11590-011-0320-4}
\BIBentrySTDinterwordspacing

\bibitem{benders2005partitioning}
J.~F. Benders, ``Partitioning procedures for solving mixed-variables programming problems,'' \emph{Computational Management Science}, vol.~2, no.~1, pp. 3--19, 2005.

\bibitem{dantzig1961decomposition}
G.~B. Dantzig and P.~Wolfe, ``The decomposition algorithm for linear programs,'' \emph{Econometrica: Journal of the Econometric Society}, pp. 767--778, 1961.

\bibitem{helsgaun2017extension}
K.~Helsgaun, ``An extension of the lin-kernighan-helsgaun tsp solver for constrained traveling salesman and vehicle routing problems,'' \emph{Roskilde: Roskilde University}, vol.~12, pp. 966--980, 2017.

\bibitem{jain1988algorithms}
A.~K. Jain and R.~C. Dubes, \emph{Algorithms for clustering data}.\hskip 1em plus 0.5em minus 0.4em\relax Prentice-Hall, Inc., 1988.

\bibitem{lucas2014ising}
A.~Lucas, ``Ising formulations of many np problems,'' \emph{Frontiers in physics}, vol.~2, p. 74887, 2014.

\bibitem{2022HybridSF}
\BIBentryALTinterwordspacing
D-Wave, ``Hybrid solvers for quadratic optimization,'' 2022. [Online]. Available: \url{https://api.semanticscholar.org/CorpusID:251834718}
\BIBentrySTDinterwordspacing

\bibitem{ruan2020quantum}
Y.~Ruan, S.~Marsh, X.~Xue, Z.~Liu, J.~Wang \emph{et~al.}, ``The quantum approximate algorithm for solving traveling salesman problem,'' \emph{Computers, Materials and Continua}, vol.~63, no.~3, pp. 1237--1247, 2020.

\bibitem{srinivasan2018efficient}
K.~Srinivasan, S.~Satyajit, B.~K. Behera, and P.~K. Panigrahi, ``Efficient quantum algorithm for solving travelling salesman problem: An ibm quantum experience,'' \emph{arXiv preprint arXiv:1805.10928}, 2018.

\bibitem{laporte2002classical}
G.~Laporte and F.~Semet, ``Classical heuristics for the capacitated vrp,'' in \emph{The vehicle routing problem}.\hskip 1em plus 0.5em minus 0.4em\relax SIAM, 2002, pp. 109--128.

\bibitem{feld2019hybrid}
S.~Feld, C.~Roch, T.~Gabor, C.~Seidel, F.~Neukart, I.~Galter, W.~Mauerer, and C.~Linnhoff-Popien, ``A hybrid solution method for the capacitated vehicle routing problem using a quantum annealer,'' \emph{Frontiers in ICT}, vol.~6, p.~13, 2019.

\bibitem{helsgaun2000effective}
K.~Helsgaun, ``An effective implementation of the lin--kernighan traveling salesman heuristic,'' \emph{European journal of operational research}, vol. 126, no.~1, pp. 106--130, 2000.

\bibitem{QiskitOptimization}
{Qiskit contributors}, ``Qiskit: An open-source framework for quantum computing,'' 2023.

\bibitem{reinelt1995tsplib95}
G.~Reinelt, ``Tsplib95,'' \emph{Interdisziplin{\"a}res Zentrum f{\"u}r Wissenschaftliches Rechnen (IWR), Heidelberg}, vol. 338, pp. 1--16, 1995.

\bibitem{lawler1966branch}
E.~L. Lawler and D.~E. Wood, ``Branch-and-bound methods: A survey,'' \emph{Operations research}, vol.~14, no.~4, pp. 699--719, 1966.

\bibitem{nabli2009overview}
H.~Nabli, ``An overview on the simplex algorithm,'' \emph{Applied Mathematics and Computation}, vol. 210, no.~2, pp. 479--489, 2009.

\bibitem{montanaro2020quantum}
A.~Montanaro, ``Quantum speedup of branch-and-bound algorithms,'' \emph{Physical Review Research}, vol.~2, no.~1, p. 013056, 2020.

\bibitem{nannicini2022fast}
G.~Nannicini, ``Fast quantum subroutines for the simplex method,'' \emph{Operations Research}, 2022.

\bibitem{karmarkar1984new}
N.~Karmarkar, ``A new polynomial-time algorithm for linear programming,'' in \emph{Proceedings of the sixteenth annual ACM symposium on Theory of computing}, 1984, pp. 302--311.

\bibitem{quetschlich2023mqt}
N.~Quetschlich, L.~Burgholzer, and R.~Wille, ``Mqt predictor: Automatic device selection with device-specific circuit compilation for quantum computing,'' \emph{arXiv preprint arXiv:2310.06889}, 2023.

\bibitem{salm2020nisq}
M.~Salm, J.~Barzen, U.~Breitenb{\"u}cher, F.~Leymann, B.~Weder, and K.~Wild, ``The nisq analyzer: automating the selection of quantum computers for quantum algorithms,'' in \emph{Symposium and Summer School on Service-Oriented Computing}.\hskip 1em plus 0.5em minus 0.4em\relax Springer, 2020, pp. 66--85.

\bibitem{chang2020integer}
C.~C. Chang, C.-C. Chen, C.~Koerber, T.~S. Humble, and J.~Ostrowski, ``Integer programming from quantum annealing and open quantum systems,'' \emph{arXiv preprint arXiv:2009.11970}, 2020.

\bibitem{johnson1996cliques}
D.~S. Johnson and M.~A. Trick, \emph{Cliques, coloring, and satisfiability: second DIMACS implementation challenge, October 11-13, 1993}.\hskip 1em plus 0.5em minus 0.4em\relax American Mathematical Soc., 1996, vol.~26.

\bibitem{uchoa2017new}
E.~Uchoa, D.~Pecin, A.~Pessoa, M.~Poggi, T.~Vidal, and A.~Subramanian, ``New benchmark instances for the capacitated vehicle routing problem,'' \emph{European Journal of Operational Research}, vol. 257, no.~3, pp. 845--858, 2017.

\bibitem{augerat1995computational}
P.~Augerat, J.~Belenguer, E.~Benavent, A.~Corber{\'a}n, D.~Naddef, and G.~Rinaldi, ``Computational results with a branch and cut code for the capacitated vehicle routing problem research report 949-m,'' \emph{Universite Joseph Fourier, Grenoble, France}, 1995.

\end{thebibliography}

\end{document}